\documentclass{PoS}

\usepackage{graphicx}

\newcommand {\bc}{\begin{center}}
\newcommand {\ec}{\end{center}}
\newcommand {\bea}{\begin{eqnarray}}
\newcommand {\eea}{\end{eqnarray}}
\newcommand {\be}{\begin{equation}}
\newcommand {\ee}{\end{equation}}
\def\R{{\mathbb R}}
\def\S{{\mathbb S}}

\title{Deconfinement transition at weak coupling in Yang-Mills 
theory on a torus}

\ShortTitle{Deconfinement transition}

\author{\speaker{Thomas Sch\"afer}\\
        Department of Physics,
        North Carolina State University, Raleigh NC 27695\\
        E-mail: \email{tmschaef@ncsu.edu}}

\abstract{We describe a weak coupling realization of the 
deconfinement transition in gauge theory compactified on 
$\R^3\times \S^1$. We consider Yang-Mills theory with a single 
Weyl fermion of mass $m$ in the adjoint representation of the 
gauge group. The fermion is subject to periodic boundary 
conditions, $\lambda(0)=\lambda(L)$, where $L$ is the size
of the circle $S_1$. This theory reduces to thermal Yang-Mills 
theory in the limit $m\to\infty$. In the limit $m\to 0$ the 
deconfinement transition can be studied using weak coupling 
methods. The analysis is based on semi-classical objects 
characterized by topological and magnetic charges. At 
leading order the relevant configurations are monopole-instantons 
and monopole-anti-monopole pairs (``bions''). We argue 
that in the $m-L$ plane the weak coupling transition is 
continuously connected to the deconfinement transition in 
pure gauge theory. }

\FullConference{31st International Symposium on Lattice Field Theory - LATTICE 2013\\
		July 29 - August 3, 2013\\
		Mainz, Germany}

\begin{document}

\section{Introduction}
\label{sec_intro}

 Finding controlled approximations to study the deconfinement 
transition in QCD, or in gauge theories related to QCD, is desirable 
for many reasons. Recently there has been some progress in this 
direction by investigating novel compactifications
\cite{Aharony:2005bq,Unsal:2010qh}. In this contribution we 
summarize recent work on gauge theory on $\R^3\times \S^1$
\cite{Poppitz:2012sw,Poppitz:2012nz} . We will argue that we can 
construct a theory that is continuously connected (as a function of a 
mass parameter) to pure gauge theory at finite temperature, and that this 
theory posses a deconfinement transition that can be studied in 
weak coupling.

\begin{figure}[t]
\bc\includegraphics[width=0.5\hsize]{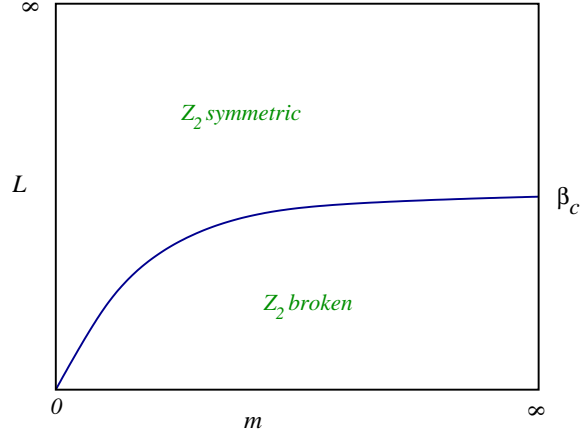}\ec
\caption{\label{fig_phase}
Schematic phase diagram of $SU(2)$ gauge theory with one 
adjoint Weyl fermion of mass $m$ compactified on $\R^3\times \S^1$.
The length of the circle is denoted by $L$. As $m\to\infty$ the 
theory has a deconfining phase transition at $L=\beta_c$, where 
$\beta_c=1/T_c$ is the inverse critical temperature of the pure 
gauge theory. }
\end{figure}

 We consider gauge theory with a single Weyl fermion in the adjoint 
representation of the gauge group. The gauge group can be any
semi-simple compact Lie group. The lagrangian is 
\be 
 {\cal L} = -\frac{1}{4g^2} F^a_{\mu\nu}F^{a\,\mu\nu}
   +\frac{i}{g^2} \lambda^a\sigma\cdot D^{ab}\lambda^b
   +\frac{m}{g^2} \lambda^a\lambda^a \, . 
\ee
Both fermions and bosons satisfy periodic boundary conditions on the 
circle, $\lambda(0)=\lambda(L)$ and $A_\mu(0)=A_\mu(L)$. A proposed 
phase diagram for this theory as a function of the compactification 
scale $L$ and the mass $m$ is shown in Fig.~\ref{fig_phase}. At $m=0$
the theory reduces to ${\cal N}=1$ SUSY Yang-Mills theory. The 
twisted partition function is equal to the Witten index, and there
is no deconfinement transition as a function of $L$. We will show 
below that for small $m$ there is a deconfinement transition at small 
$L$. In $SU(N)$ gauge theories this transition is characterized
by the breaking of $Z_{N}$ symmetry. As $m\to\infty$ the theory
reduces to thermal pure gauge theory which is known to have a 
deconfinement phase transition at $L=\beta_c=1/T_c$. This transition
is second order for $SU(2)$ gauge theory, and first order for $SU(N
\geq 3)$ or other higher rank gauge groups. Fig.~\ref{fig_phase}
shows the minimal phase diagram consistent with these facts. It 
is possible that there are additional transitions at intermediate
$m$ that are not associated with a change of symmetry. It is also 
possible that the slope of the transition line is not positive 
everywhere. This would not invalidate the picture presented here, 
but it would make extrapolation from small $m$ to large $m$ more 
difficult. Both of these possibilities can be investigated using 
lattice simulations.

\section{Weak coupling calculation}
\subsection{Effective theory for small $\S^1$}

 In this section we will focus on $SU(2)$ gauge theory. Classical vacua
of the theory are labeled by the Polyakov line
\be 
 \Omega = \exp\left( i\int A_4 dx_4 \right)\, . 
\ee
The Polyakov line can be diagonalized, $\Omega={\rm diag}
(e^{i\Delta\theta/2},e^{-i\Delta\theta/2})$. At a generic point on 
the moduli space $\Delta\theta\neq 0$ and the Polyakov line acts as
a Higgs field that breaks the gauge symmetry to its abelian subgroup,
$SU(2)\to U(1)$. We can construct an effective theory that describes
the light fields in the limit that the size $L$ of circle $\S^1$ is 
small. 

 In this limit we can focus on the lowest Kaluza-Klein modes, and 
the effective lagrangian involves three dimensional fields. There 
are two light bosonic fields. One is the massless photon associated
with the unbroken $U(1)$ symmetry. We describe this field using 
the dual photon $\epsilon_{ijk}\partial_k\sigma=F_{ij}$. We will 
see that near $m=0$ the potential for $\Delta\theta$ is almost 
flat and the second light field is associated with fluctuations
of the holonomy. We define $b=4\pi\Delta\theta/g^2$. Finally, 
there is one light fermionic field $\lambda^\alpha$ which is 
associated with the abelian subgroup. For $m=0$, these three fields
can be written in terms of a chiral superfield $B=b+i\sigma + 
\sqrt{2}\theta^\alpha\lambda_\alpha$. The effective lagrangian
for the bosonic fields is 
\be
{\cal L} = \frac{g^2}{32\pi^2L}\left[
(\partial_i b)^2+(\partial_i\sigma)^2 \right] + V(\sigma,b)\, , 
\ee
where we have determined the kinetic terms at leading order in
perturbation theory.

\subsection{Perturbative effects}

 The scalar potential $V(\sigma,b)$ has an expansion of the form
\be
V = \sum_n g^{n}V^n_0
         + \sum_n g^{n}e^{-\frac{c_0}{g^2}}V^n_1
         + \sum_n g^{n}e^{-\frac{2c_0}{g^2}}V^n_2
         + \ldots \, ,
\ee
where $V^n_0$ is related to perturbative effects and $V^n_k$ is 
determined by semi-classical configuration with action $S=kc_0/g^2$.
At one-loop order the perturbative part of the potential was computed
by Gross, Pisarski and Yaffe \cite{Gross:1980br}. In ${\cal N}=1$ 
SUSY YM theory the potential vanishes because bosonic and fermionic
contributions cancel. If the mass of the fermion is not zero then 
the cancellation is not exact. We find
\be
\label{V_gpy}
V =  - \frac{m^2}{2 \pi^2L^2}
      \sum_{n=1}^{\infty}   \frac{1}{n^2} \,  
      \left|{\rm tr}\, \Omega^n \right|^2 
  =  - \frac{m^2}{L^2} B_2 \left( \frac{\Delta\theta}{2\pi} \right)\, , 
\ee
where $B_2$ is the second Bernoulli polynomial. There is no potential
for the dual photon. The potential for the holonomy has a minimum at 
$\Delta\theta=0,2\pi$, which corresponds to the $Z_2$ broken phase. 
The center symmetric point $\Delta\theta=\pi$ is a local maximum of 
the potential. 

\subsection{Non-perturbative effects: SUSY Yang-Mills theory}

\begin{figure}[t]
\bc\includegraphics[width=0.7\hsize]{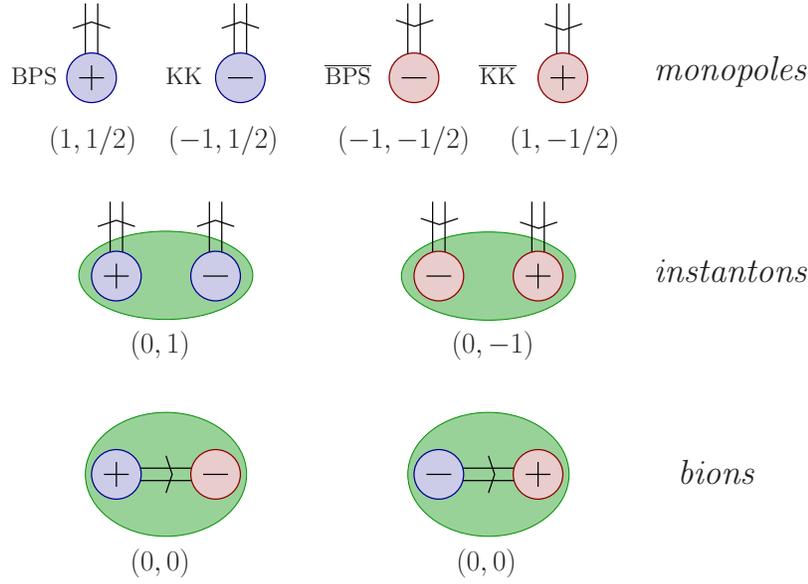}\ec
\caption{\label{fig_top}
Topological objects in $SU(2)$ gauge theory on $\R^3\times \S^1$.
The objects are labeled by $(Q_M,Q_{top})$ for the center-symmetric 
holonomy. Lines denoted fermionic zero modes, and the arrows 
point from $\lambda\lambda$ to $\bar\lambda\bar\lambda$ vertices.}
\end{figure}

 For $m=0$ the potential for $\Delta\theta$ vanishes to all orders
in perturbation theory. This implies that exponentially small corrections
that arise from topological objects are important even at small coupling. 
Semiclassical objects on $\R^3\times \S^1$ can be classified by the asymptotic 
value of the holonomy $\Omega$ and by their topological and 
magnetic charges \cite{Gross:1980br}
\be
\left(Q_M,Q_{top}\right) = 
\left( \frac{1}{4\pi}\int_{S_2} B\cdot d\Sigma , 
    \frac{1}{32\pi^2}\int_{\R^3\times \S^1} F_{\mu\nu}^a
                              \tilde{F}^{\mu\nu\,a} \right)\, . 
\ee
Periodic instantons (calorons) are topological objects with 
$Q_{top}=k$ ($k\in Z$) and magnetic charge zero. Monopole-instantons, 
also known as dyons, are magnetically charged objects with fractional 
topological charge. Monopole-instantons come in two types, which we will 
refer to as BPS and KK monopole-instantons \cite{Lee:1997vp,Kraan:1998sn}. 
Instantons can be viewed as bound states of BPS and KK monopoles. In 
particular, the magnetic charges of the two types of monopoles are 
opposite, and their topological charges add to an integer, see 
Fig.~\ref{fig_top}.

 The coupling of the elementary BPS and KK monopoles to the low
energy fields is given by 
\bea 
{\cal M}_{1} =       e^{-b+i\sigma}  \lambda\lambda,  && 
{\cal M}_{2} = \eta  e^{+ b-i\sigma} \lambda\lambda,  \\
\overline{\cal M}_{1} =     e^{-b-i\sigma} \bar\lambda\bar\lambda, &&
\overline{\cal M}_{2} =\eta e^{+b+i \sigma} \bar\lambda\bar\lambda, 
\eea
where $\eta=\exp(-2S_0)$ with $S_0=4\pi^2/g^2$ and we have suppressed
overall numerical factors. Monopole-instantons carry fermionic zero 
modes and do not contribute to the bosonic potential. In the SUSY 
Yang-Mills case fermion zero modes are lifted by the integral over 
Grasssmann parameters, and monopole-instantons give a non-zero 
superpotential. We find \cite{Davies:2000nw}
\be
{\cal W} = \frac{M_{PV}^3L}{g^2}\left( e^{-B}+e^{-2S_0}e^B\right)\, ,
\ee
where $M_{PV}$ is a Pauli-Villars mass parameter. This is the 
Affleck-Dine-Seiberg superpotential, which was originally determined 
by different methods \cite{Affleck:1983mk}. The scalar potential is
\be
\label{V_susy}
 V(b,\sigma) \sim \left|\frac{\partial {\cal W}}{\partial B}\right|^2
 \sim \frac{M_{PV}^6L^3e^{-2S_0}}{g^6}\left[
  \cosh\left(\frac{8\pi}{g^2}\left(\Delta \theta-\pi\right)\right) 
 -\cos(2\sigma)\right]\, . 
\ee
We observe that the potential has a minimum at the center-symmetric
point $\Delta\theta=\pi$, and that there is a mass gap for the dual
photon. This means that for $m=0$ the theory is in the confined
phase for all $L$.

\subsection{Non-perturbative effects: Non-zero mass case}

 In this section we show how to rederive this result without using 
supersymmetry, and then extend the calculation to $m\neq 0$. The relation
$V\sim |\partial{\cal W}/\partial B|^2$ implies that  the monopole 
contribution to the superpotential corresponds to a monopole-anti-monopole 
contribution to the scalar potential. In particular, the potential for 
the dual photons is generated by magnetic ``bions'' $[{\cal M}_1
\overline{\cal M}_2]$ and $[{\cal M}_2\overline{\cal M}_1]$, and the 
potential for the holonomy is generated by neutral ``bions'' 
$[{\cal M}_1\overline{\cal M}_1]$ and $[{\cal M}_2\overline{\cal M}_2]$ 
\cite{Unsal:2007jx}.

 Calculating the contribution of neutral bions is subtle because
the topological charge is zero and there is no barrier between the 
semi-classical contribution and the perturbative vacuum. The amplitude
is of the form 
\be
{\cal A}_{[{\cal M}_1\overline{\cal M}_1]} \sim e^{-2b} 
\int d^3r\, e^{-S_{12}(r)}, \hspace{0.05\hsize}
  S_{12}(r) = -2 \frac{4\pi L}{g^2 r}  + 4\log(r)\, 
\ee
where $d^3r$ is the integral over the monopole separation.
The first term in $S_{12}$ is the scalar attraction between
the monopoles, and the second term is due to approximate fermion 
zero modes. The integral over $r$ diverges at small $r$. In 
\cite{Poppitz:2012sw} we show how to compute the amplitude by 
analytic continuation in $g^2$ \cite{Poppitz:2012sw}.
This method was introduced by Bogomolny and Zinn-Justin (BZJ)
in the context of instanton-anti-instanton calculations in 
quantum mechanics. We show that the total contribution from 
neutral bions is given by
\be
 V(b,\sigma) \sim \frac{M_{PV}^6L^3e^{-2S_0}}{g^6}
  \cosh\left(\frac{8\pi}{g^2}\left(\Delta \theta-\pi\right)\right) \, , 
\ee
in agreement with the calculation based on the superpotential.

 Once we know how to compute the potential without supersymmetry
it is straightforward to extend the result to $m\neq 0$. There 
are three contributions: 1) The perturbative potential given in 
equ.~(\ref{V_gpy}), 2) the potential from neutral and charged 
bions, 3) a contribution from monopoles in which the fermion 
zero mode is lifted by the mass term. We find \cite{Poppitz:2012sw}
\be
\tilde{V} = 
  \cosh 2 b^\prime - \cos 2 \sigma  +
   \frac{\tilde{m}}{2\tilde{L}^2} \cos\sigma 
   \left( \cosh b^\prime
       - \frac{b^\prime \sinh b^\prime}{3\log\tilde{L}^{-1}}\right) 
 - \frac{1}{1728\log^3\tilde{L}^{-1}}  
   \left({\tilde{m}  \over  \tilde{L}^2}\right)^2 
     (b^\prime)^2 \, , 
\ee
where we have introduced dimensionless variables $b'=b-4\pi^2/g^2$, 
$\tilde{L}=\Lambda L$, and $\tilde{m}=m/\Lambda$. $\Lambda$ is the 
scale parameter, and $\tilde{V}$ is a dimensionless potential, see 
equ.~(2.35) in \cite{Poppitz:2012sw}. The competition between the 
center stabilizing bions and the center de-stabilizing monopoles and 
perturbative terms leads to a phase transition at $L_c=\Lambda^{-1}
(\tilde{m}/8)^{1/2}$, consistent with the phase diagram shown in 
Fig.~\ref{fig_phase}.

\section{Extension to other gauge groups and outlook}

 In \cite{Poppitz:2012nz} we show how to extend this analysis to higher 
rank gauge groups. We consider both gauge groups with and without a 
non-trivial center. If the case of gauge groups with a trivial center, like 
$G_2$, the deconfinement transition is not associated with a change of 
symmetry. For a general gauge group of rank $r$ there are $r$-1 fundamental 
BPS and one fundamental KK monopole-instanton \cite{Davies:2000nw}. The 
monopole and bion induced potentials can be expressed in terms of the 
roots of the Lie algebra. 

\begin{figure}[t]
\bc\includegraphics[width=0.41\hsize]{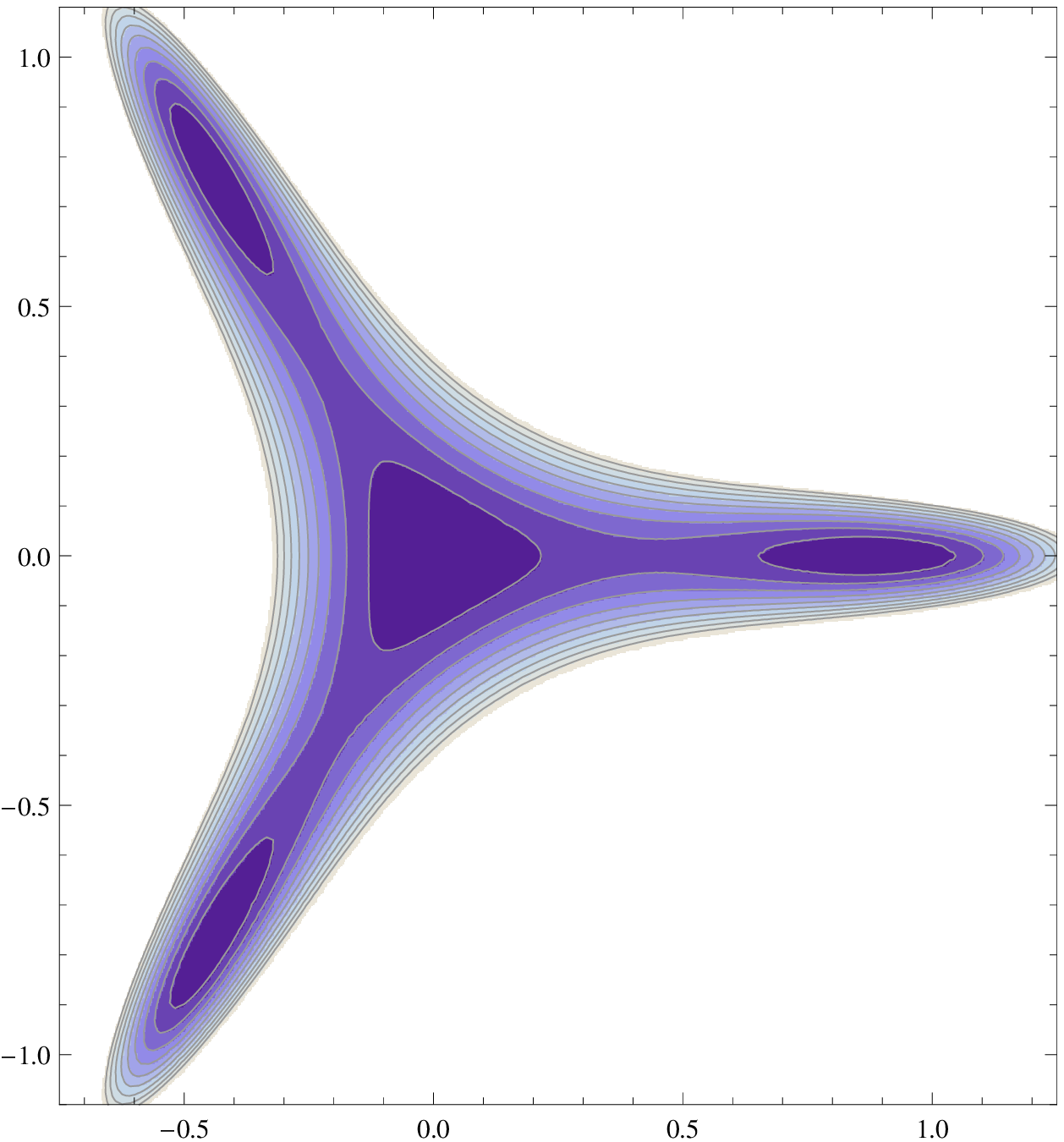}
\hspace*{0.075\hsize}
\includegraphics[width=0.45\hsize]{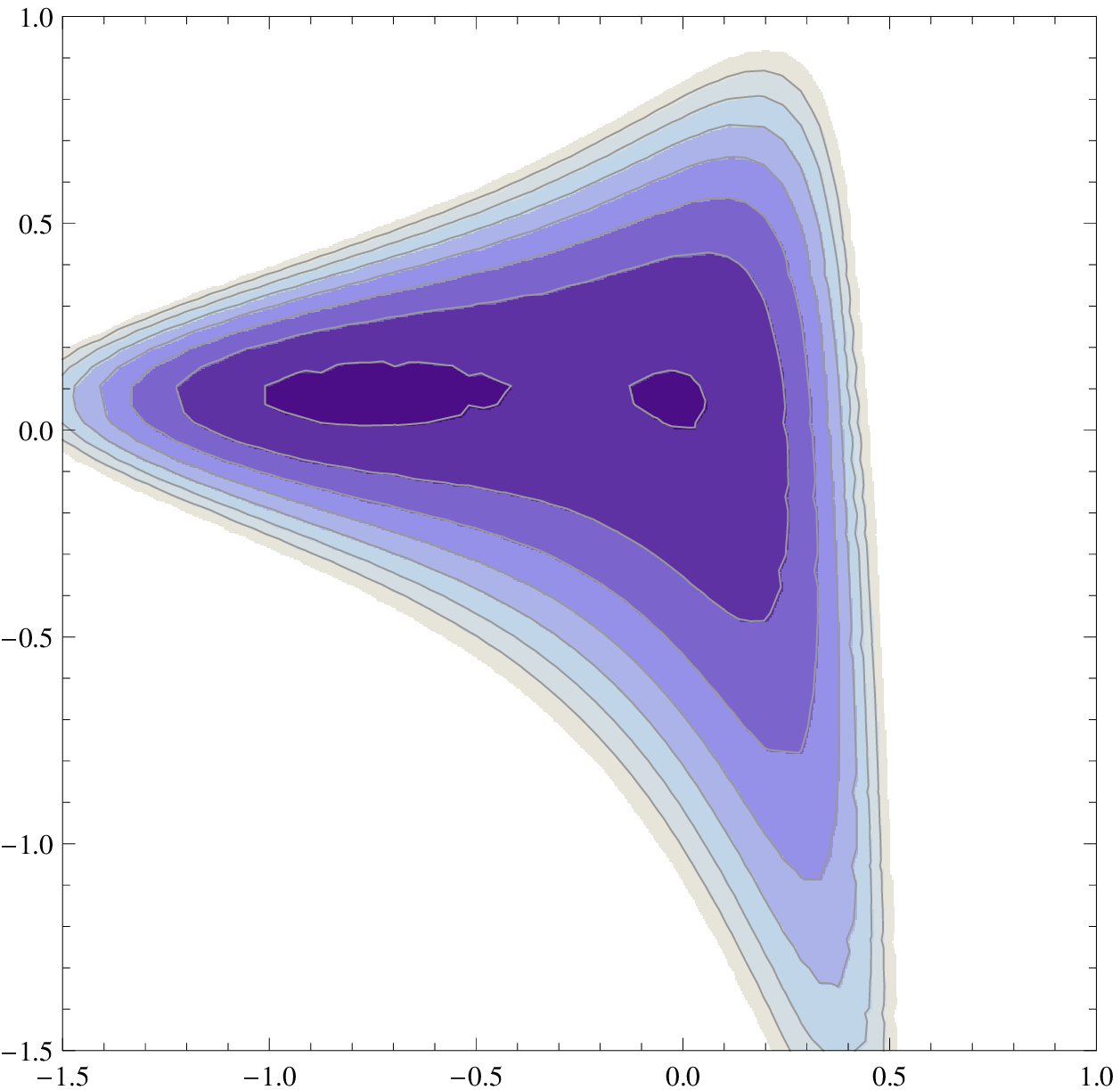}\ec
\caption{\label{fig_SU3_G2}
Contour plots of the effective potential for the holonomy in $SU(3)$
(left panel) and $G_2$ gauge theory (right panel). The potentials are 
shown at the critical length $L_c$ corresponding to the first order 
deconfinement transition. The holonomy is written as $\Omega=\exp(i
\vec{H}\cdot(\vec{b}_0+\vec{b}))$ and plotted as a function of $b_{1,2}$.
See \cite{Poppitz:2012nz} for the definition of the Cartan vector $\vec{H}$
and the center symmetric holonomy $\vec{b}_0$.}
\end{figure}

 The phase diagram for higher rank gauge groups has the same structure 
as the $SU(2)$ phase diagram shown in Fig.~\ref{fig_phase}, except that 
the phase transition is first order. In Fig.~\ref{fig_SU3_G2} we show 
contour plots of the potential for the holonomy at the deconfinement 
transition in the case of $SU(3)$ and $G_2$ gauge theory. Both transitions
are clearly first order, but in the case of $G_2$ the holonomy is 
non-vanishing in both phases. There are a variety of issues that can 
be studied:

\begin{itemize}
\item The large $N_c$ limit is smooth provided the mass of the lightest
higgsed gluon, $m_w\sim 2\pi/(N_cL)$, is kept fixed as $N_c\to\infty$. 
The effective potential has multiple branches labeled by $k=0,\ldots,
N_c-1$, in agreement with Witten's arguments 
\cite{Witten:1980sp,Witten:1998uka}.

\item We can compute the shift in $L_c$ due to a non-zero theta term. 
We observe that the critical $T_c\sim L_c^{-1}$ is reduced 
\cite{Anber:2013sga}, in agreement with lattice calculations reported
in \cite{D'Elia:2012vv}. 

\item We have studied the distribution of the eigenvalues of the 
Polyakov line in the confined and deconfined phases. We observe the
expected eigenvalue repulsion in the confined phase, and clustering 
in the deconfined phase. In the case of $G_2$ we observe that the 
Polyakov line jumps from a slightly negative value below $T_c$ to 
a positive value above $T_c$. This behavior was also seen in lattice
calculations \cite{Holland:2003jy}.

\end{itemize}

 Recent work has also begun to address the role of fermions in the 
fundamental representation. For large quark masses one finds the 
expected effects due to explicit breaking of the center symmetry
\cite{Poppitz:2013zqa}. For small quark masses the theory flows
to strong coupling, and a dual description is required.

 Acknowledgments: I would like to hank my collaborators Erich 
Poppitz and Mithat \"Unsal. This work was supported by the US 
Department of Energy grant DE-FG02-03ER41260.

\end{document}